\documentclass[a4paper, 11pt]{article}
\usepackage{mathrsfs}
\usepackage{amsmath}
\usepackage{amsfonts}
\usepackage{amssymb}
\usepackage[utf8]{inputenc}
\usepackage{appendix}
\usepackage[english]{babel}    
\usepackage{graphicx}         
\usepackage{float}          
\usepackage[usenames]{color}  
\usepackage{anysize}		
\marginsize{2cm}{2cm}{2cm}{2cm}  
\usepackage{authblk}
\usepackage{url}

\title{Robustness of the public transport network against attacks on its routes}

\author[1,2]{Tomás Cicchini} 
\author[1,4]{Inés Caridi} 
\author[3,4]{Leonardo Ermann}

\affil[1]{CONICET-Universidad de Buenos Aires, Facultad de Ciencias Exactas y Naturales, Instituto de Cálculo (IC)}

\affil[2]{Universidad de Buenos Aires, Facultad de Ciencias Exactas y Naturales, Departamento de Física}

\affil[3]{Dto FT, GIyA, Comisión Nacional de Energía Atómica} 

\affil[4]{CONICET}

\date{}
\begin{document}
\maketitle

\begin{abstract}
We investigate the robustness of Public Transport Networks (PTNs) when subjected to route attacks, focusing specifically on public bus lines. Such attacks, mirroring real-world scenarios, offer insight into the multifaceted dynamics of cities. Our study delves into the consequences of systematically removing entire routes based on strategies that use centrality measures.
We evaluate the network's robustness by analyzing the sizes of fragmented networks, focusing on the largest components and derived metrics. To assess the efficacy of various attack strategies, we employ them on both a synthetic PTN model and a real-world example, specifically the Buenos Aires Metropolitan Area in Argentina.
We examine these strategies and contrast them with random, and one-step most and least harmful procedures. Our findings indicate that \textit{betweenness}-based attacks and the one-step most (\textit{maximal}) harmful procedure emerge as the most effective attack strategies. Remarkably, the \textit{betweenness} strategy partitions the network into components of similar sizes, whereas alternative approaches yield one dominant and several minor components.
\end{abstract}




\section{Introduction}

During the last decades, there has been a growing focus on central aspects of urban infrastructure and its functioning within the framework of a nascent science of cities \cite{Batty2017}. This surge in interest has been propelled by the utilization of spatial networks and the accessibility of public data alongside Geographical Information System (GIS) tools. Moreover, the perspective of considering cities as complex systems, as their overall properties are emergent and are not simple summations of their parts, is increasingly relevant \cite{Ortmanid2020}. Thus, many efforts seek to understand the social aspects underlying the dynamic change of cities driven by human mobility \cite{Batty2008, Xu2023}, design innovative solutions for city services \cite{KUMAR2020119281}, and plan improvements that impact the mobility of inhabitants \cite{Perez-Mendez2021, Yinger2021, Cervero2017}.

An essential infrastructure facilitating the daily movement of people within a city is the public transportation system, which comprises various routes or lines of different modes such as buses, trains, metros, and other public conveyance means. The Public Transport Network (PTN) integrates data from the public transportation system with spatial information about cities \cite{VonFerber2007, VonFerber2009, Bahman_trainvulnerability}. However, there is no singular method for representing the PTN solely based on information regarding a specific mode, such as buses. Instead, various representations exist to depict public transportation information within specific networks, including $\mathbb{L}$-space, $\mathbb{L'}$-space, $\mathbb{C}$-space, and $\mathbb{P}$-space \cite{Barthelemy2011, Pan2023}. Each of these representations is tailored to serve distinct analytical purposes or to study particular processes of interest in the transportation system, providing insights into different facets of transportation operations. In this study, our focus lies on examining the impact of route removal on the connectivity of the transportation system.

Each route, such as a bus line, consists of a sequential arrangement of stations spanning from an originating station to a terminal one. Passengers have the option to board or disembark at each station along the route. Our motivation in this study is to analyze scenarios wherein an entire bus line discontinues operation either due to a failure or attack, or based on a particular decision of the line's company management or a strike by its employees, disrupting then the route as a whole.

In many interconnected complex systems across various domains such as biological, social, economic, and technological realms, public transportation networks can experience both intentional and unintentional failures, impacting individual system elements as well as the system as a whole \cite{artime2024}. Unintentional shortcomings typically manifest as minor random disruptions within the network, such as stations or lines being temporarily disabled. Sometimes, these failures are part of the dynamics of networks, making the study of cascading system crashes essential for understanding systems evolution, as discussed in \cite{LICascading, XiaoCrash}.  On the other hand, intentional failures, often termed attacks, aim to disrupt the entire network's operation. These attacks commonly target nodes, links, or sets of nodes with the most significant potential to destabilize the network. The exploration of how diverse networks amplify failures at the systemic level, along with their stability and recovery, traverses numerous disciplines and has been investigated in networks beyond transportation, including biological, ecological, financial, and social networks \cite{MARTINEZPASTOR2022100570}. This subject matter is closely related to the robustness and resilience of networks and systems \cite{artime2024}, garnering considerable attention due to its significance in identifying and mitigating problems. Various solutions have been proposed, such as interventions to mitigate wildfires and extreme events \cite{ARANGO_wildfire}, the development of early warning systems for different disruptive scenarios, the identification of critical and vulnerable elements within the system \cite{MARTINEZPASTOR2022100570, Wang2023_1, Wang2023_2}, and the proposal of improved networks to prevent future failures \cite{Wang2023_3}. 

In the context of PTN, failures can potentially  disrupt the daily lives of passengers and all city residents, leading to travel delays for passengers and impacting the city's broader transportation system. Hence, it is crucial to understand the susceptibility of PTN to failures or attacks to anticipate critical collapses \cite{Scheffer}, plan future improvements in the city's transportation infrastructure, and effectively respond to eventual collapses following a failure. The robustness of PTN has been examined across various cities, considering different attack strategies targeting nodes and links \cite{Candelieri2019,RODRIGUEZNUNEZ201450,modiri}.

The central question guiding this research is to understand which type of failure or attack on the bus public transportation network yields the most significant impact on the city. It is worth noting that this approach differs from previous studies, which primarily focused on directed attacks targeting specific nodes or links within the network \cite{VonFerber2007,vonFerber2012,BERCHE_2012}. We operate under the assumption that each transport line, like a bus line, functions as a cohesive unit integrating a set of employees, a fixed organization, and logistical arrangements. As a result, the interruption extends from the starting station to the end station along the route. This paper compares and analyzes the effects of various strategies for selecting the removed line from the PTN based on network topological properties. Our objective is to evaluate which route attack strategy inflicts more significant damage to the network and quantify its impact. The proposed methodology aligns with the broader theory of bond (or edge) percolation. For instance, in the realm of PTN analysis, prior works such as \cite{Berche2009,Berche2010PublicTN,zhang} concentrate on attacking edges based on their centrality measures and delineate the effects of such attacks. However, when a route is attacked, instead of removing a single edge at each step from $\mathbb{L}$-space, all its set of consecutive edges are eliminated from the $\mathbb{L'}$-space. Indeed, although there exist several works investigating the impact of removing straight rigid rods \cite{Ramirez2015,Ramirez2018} or $k^2$-mers \cite{Ramirez2019} within the framework of regular lattices, the exploration of removing complex paths in real networks remains relatively unexplored.

To analyze general cities and compute statistics of PTN, we will generate a synthetic PTN based on the model proposed in \cite{VonFerber2007}. These synthetic networks emulate topological properties similar to those observed in large cities \cite{VonFerber2009}, incorporating a random component to allow for multiple realizations.

Moreover, we will conduct a similar analysis on the PTN network of a Latin American city, specifically focusing on public bus transportation. To do so, we will utilize open data from \cite{OpenStreetMap} to examine 489 bus lines within the Metropolitan Area of Buenos Aires (AMBA), Argentina. This region spans approximately 13,285 square kilometers  \cite{ambaSup} and serves around four million people daily \cite{ambaUso}.

The paper's organization is structured as follows: Section \ref{methodology} outlines the PTN methodology, details the route removal strategies for implementing attacks, discusses the metrics used to quantify the impacts of the attacks, and provides a description of the synthetic PTN employed. Section \ref{results} presents the results of the attacks on the synthetic PTN. In Section \ref{real PTN}, the same methodology is applied to the city of Buenos Aires. Section \ref{discussion} covers the discussion and future research directions.
\section{Methodology} \label{methodology}

\subsection{Public transport network and its representations}

In this study, our goal is to examine the robustness of Public Transport Networks (PTNs) within metropolitan areas. We define a PTN as the entire public transport system, encompassing the infrastructure (stops, stations, and rails, where applicable) as well as the routes of various bus, subway, and train lines that serve the mobility needs of inhabitants. As established in the literature \cite{Barthelemy2011}, several methods exist to represent such a structure as a network.

The simplest and most common representation is known as the $\mathbb{L}$-space, where nodes correspond to stops and stations, and an edge exists between two nodes if they are consecutive stops on at least one route. A variation of this approach, known as the $\mathbb{L'}$-space, employs a distinct edge type for each route connecting consecutive nodes. While the $\mathbb{L}$-space is a simple network, the $\mathbb{L'}$-space is a multi-edge network; however, both are considered spatial networks due to the geographical embedding of the nodes.

Alternatively, we can consider a network where nodes represent the different routes of the PTN, and edges indicate that two routes share at least one station. This representation is called the $\mathbb{C}$-space and is a non-spatial simple network.

\begin{figure}[H]
    \centering
    \includegraphics[width=1.\textwidth]{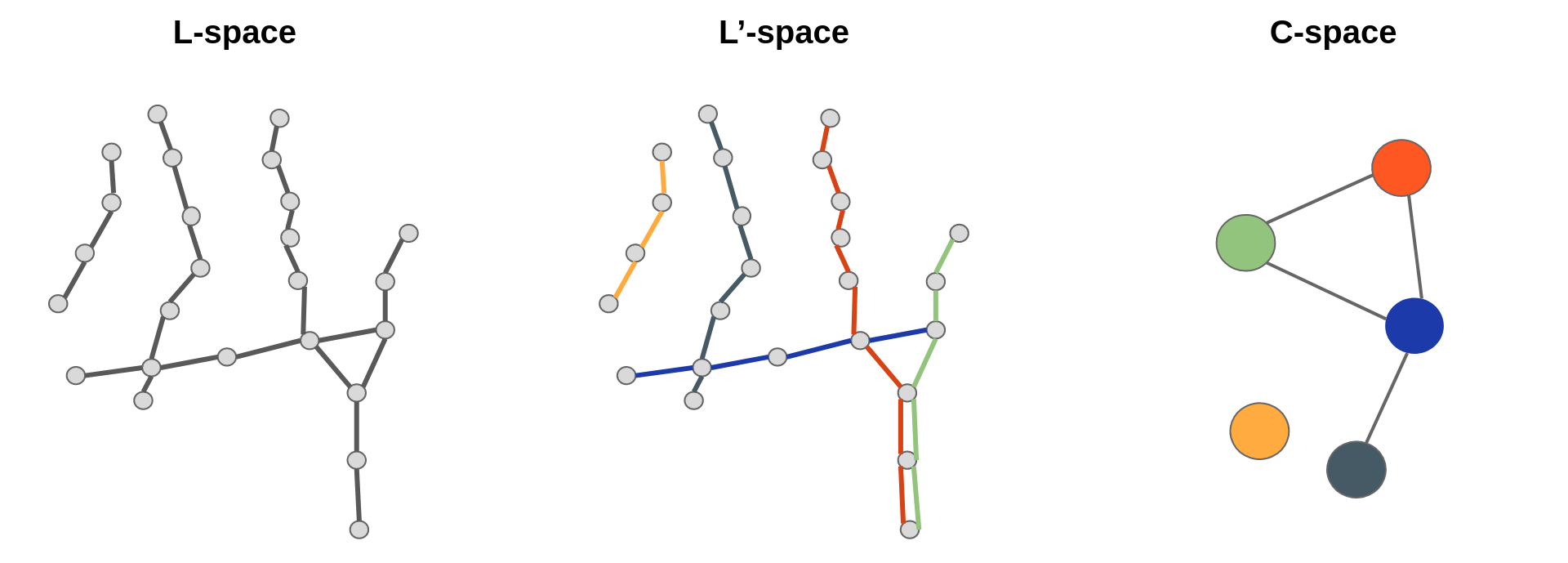}
    \caption{Example of a PTN and their three different representations, $\mathbb{L}$, $\mathbb{L}$' and $\mathbb{C}$-spaces.}
    \label{fig1}
\end{figure}

Furthermore, the size of PTN is denoted by $N$ in $\mathbb{L}$ and $\mathbb{L'}$ spaces, and $R$ in $\mathbb{C}$-space. It's important to note that in $t$ sequential route attacks, $N$ remains constant, while the fraction of attacked routes $t/R=\Phi \in [0,1]$ defines the size of the $\mathbb{C}$-space as $R-t=R(1-\Phi)$.
Figure \ref{fig1} shows an example of a PTN in $\mathbb{L}$-space, $\mathbb{L}$'-space and $\mathbb{C}$-space.

\subsection{Robustness metrics}

To quantify and measure the robustness of  the PTN against sequential removal of routes in  $\mathbb{L}$-space and $\mathbb{L}$'-space, we focus primarily on $S_1$, which represents the size of the largest connected component, also known as the \textit{giant component}. This metric is widely employed to analyze percolation phenomena and network damage \cite{artime2024, schneider2011}. However, as routes are successively removed from the PTN, additional isolated and connected components may emerge.

In order to comprehensively understand and quantify how the absence of routes impacts the PTN, we also investigate the sizes of the other connected components. Here, $S_i$ denotes the size of the $i$-th component, with $i$ ranging from $1$ to $\Omega_S$. Therefore, $\Omega_S$ represents the total number of components, satisfying the condition $N\geq S_1\geq S_2\geq\ldots\geq S_{\Omega_S}\geq1$, with $\sum_{i=1}^{\Omega_S}S_i=N$.

To gain insight into the distribution of the number of nodes in each component, it is beneficial to define the normalized entropy of components size distribution $\Sigma_S$ as follows:
\begin{equation}
    \Sigma_S = \left\{ \begin{array}{crr} 0 & \text{if} & \Omega_S=1\\ -\frac{1}{\log \Omega_S}\sum_{i=1}^{\Omega_S} p_i\log (p_i) & \text{if} & \Omega_S> 1 \end{array} \right.
\end{equation}
where $p_i$ is the fraction of nodes in the $i$-th component  $p_i=S_i/N$ and then resulting $0\leq\Sigma_S\leq1$.

To quantify the global robustness in a single metric,  we track the component sizes throughout the entire sequence of route attacks (attack strategy). 
The global robustness of each component is defined as follows:
\begin{equation}
\mathcal{R}_i = \frac{1}{NR}\sum_{t=0}^{R} S_i(\Phi) 
\end{equation}
where $\Phi=t/R$, $\Phi$ represents the ratio of attacked lines ($t$) to the initial number of lines ($R$). Here, $\mathcal{R}_i$ denotes the area below the curve $S_i(\Phi)$ between 0 and 1, which converges to the integral $\int_0^1 S_i(\Phi)d\Phi/N$ for very large number of routes $R$. By definition, $\mathcal{R}_i$ falls within the interval [0,1], and $\mathcal{R}_i > \mathcal{R}_j$ for $i < j$. A similar definition to $\mathcal{R}_1$ for node attacks, was introduced in \cite{schneider2011} as the "Unique Robustness Measure" for studying robustness under node attacks. We will analyze the cases of $\mathcal{R}_1$ and $\mathcal{R}_2$, which measure the global robustness of the giant and the second component, respectively.
The global robustness of the giant component, $\mathcal{R}_1$, can be interpreted as follows: when $\mathcal{R}_1 \approx 0$, the network is highly vulnerable to the attack. Even a small fraction of routes removed ($\Phi \approx 0$) leads to the collapse of the giant component to zero. Conversely, in the case of $\mathcal{R}_1 \approx 1$, it means a network highly resilient against such attacks. This implies that almost all existing routes ($\Phi \approx 1$) must be removed for the giant component to decrease in size.
It is worth mentioning that we will compute all metrics in $\mathbb{L}$'-space for convenience, but they are the same as those computed in $\mathbb{L}$-space since metrics are derived from the component sizes. Therefore, the analysis and interpretation of the global robustness metrics hold true for both $\mathbb{L}$-space and $\mathbb{L}$'-space.

\subsection{Attack strategies}

This work aims to investigate the robustness of the PTN under various sequential attacks, based on $\mathbb{C}$-space. The order in which each route is removed is crucial and can be regarded as an attack strategy or procedure.

We employ well-known standard centrality measures in complex networks \cite{Newman} to define sequential removal strategies, including: \textit{degree},  \textit{closeness}, \textit{PageRank}, and  \textit{betweenness} (for more details, see \ref{centrality measures}).
The removal procedure given by the largest centrality route in $\mathbb{C}$-space is dynamically updated as follows: 

\begin{enumerate}
    \item \emph{target route selection}: define the route to attack based on the  centrality measures in $\mathbb{C}$-space.
    \item \emph{attack}: remove the selected route and recalculate both $\mathbb{L}'$-space and $\mathbb{C}$-space.
    \item \emph{metric calculations}: calculate metrics over $\mathbb{L}'$-space.
    \item Repeat steps 1, 2 and 3 until the $R$ routes are removed ($\Phi=1$).
\end{enumerate}

\begin{figure}[H]
    \centering
    \includegraphics[width=1.\textwidth]{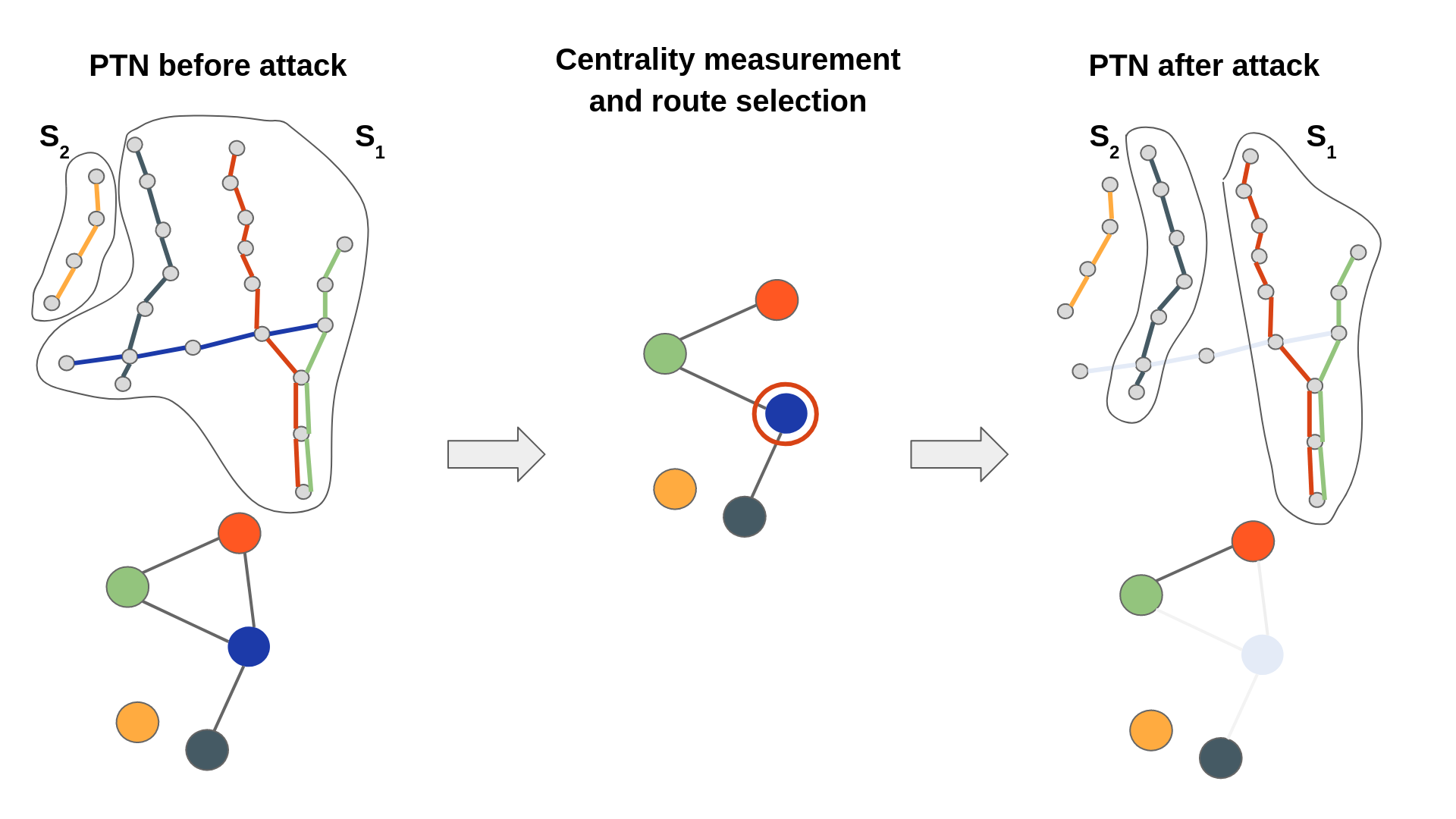}
    \caption{Schematic representation of the removal 
    procedure following the strategy of \textit{degree} centrality 
    measure for $N=25$ and $R=5$. Given a certain PTN, a centrality measurement is performed and the most important route is attacked, i.e., all its edge are removed from $\mathbb{L}$'-space and the node
    is removed from the $\mathbb{C}$-space.}
    \label{fig2}
\end{figure}

This procedure is illustrated in Figure \ref{fig2}  for \textit{degree} centrality in an example of PTN with $N=25$ and $R=5$.
The blue route has the highest \textit{degree} centrality value in  $\mathbb{C}$-space and is then removed.
Before the removal of the blue route, the giant component  has a size of $S_1=21$, which decreases to $S_1=12$ after removal.
The number of components changes from $\Omega_S=2$ to $\Omega_S=5$,  with $S_2=4$ before removal to $S_2=7$, $S_3=4$, $S_4=S_5=1$ after removal. 
Therefore, the normalized entropy  changes from $\Sigma_S\simeq0.63$ to $\Sigma_S\simeq0.78$.
 
In addition to the four different PTN attack procedures mentioned, we will also analyze the random removal of routes (representing unintentional failures), as well as two new strategies referred to as the \textit{maximal} and \textit{minimal} strategies.

The \textit{maximal} and \textit{minimal} procedures are defined as follows: starting from a $\mathbb{L'}$-space, we compute which route causes the maximum and minimum damage to the giant component size $S_1$, respectively. For instance, in the network of the left panel of Figure \ref{fig2}, the giant component size ($S_1$) decreases by $0$, $2$, $5$, $6$, and $9$ when we remove the yellow, green, red, dark gray, and blue routes, respectively. Therefore, in the \textit{maximal} and \textit{minimal} strategies, the blue and yellow routes will be removed, respectively. These procedures are repeated at each step, similar to centrality measures.

Despite previous global analysis, such as the one in \cite{DENG201674}, it is important to notice that these strategies compute the impact of only one removed route in each step. Therefore, applying the \textit{maximal} (or \textit{minimal}) strategy twice does not guarantee that these two attacked routes will maximize (or minimize) the decrease of $S_1$. In this sense, we cataloged the \textit{minimal} and \textit{maximal} attacks as one-step strategies, while in \cite{DENG201674} the optimal attack is found globally. For more details of the proposed strategies, see \ref{centrality measures}.

\subsection{Centrality measures}\label{centrality measures}

As mentioned, we select the most central route at each step based on various topological measures to determine which route in the $\mathbb{C}$-space should be removed. The simplest measure we use is the \textit{degree} centrality $k_i$, which characterizes the number of neighbors of node $i$. In the $\mathbb{C}$-space, $k_i$ represents the number of other routes that share at least one stop with route $i$.

Another centrality measure we utilize is the \textit{closeness} centrality. This measure is defined by considering the distance over the network between nodes $i$ and $j$, denoted by $d_{ij}$, which is the number of edges between them. The \textit{closeness} centrality $c_i$ of a node $i$ is then given by:
\begin{equation}
    c_i = \sum_{j \neq i} \frac{1}{d_{ij}}
\end{equation}
where $d_{ij} \to \infty$ when there is no path between nodes $i$ and $j$. Essentially, this measure quantifies the proximity of a node to all other nodes in the network. In the context of the PTN, \textit{closeness} centrality can be interpreted as a proxy for the average number of transfers a passenger would need to make from line $i$ to the other lines in the system.

Closely related to the former measure, the \textit{betweenness} centrality is defined as:
\begin{equation}
    b_i = \sum_{j\neq k \neq i \in N}\frac{\sigma^{i}_{jk}}{\sigma_{jk}}
\end{equation}
where  $\sigma_{jk}$ represents the total number of shortest paths between nodes $j$ and $k$, while $\sigma_{jk}^i$ represents the total number of shortest paths between $j$ and $k$ passing through node $i$.
This measure quantifies whether a particular route acts as a bridge between others or if it is expendable. In other words, it indicates the extent to which a route facilitates the flow of passengers between different parts of the network.

\textit{PageRank} centrality, denoted as $p_i$, is a measure of the importance or influence of a node in a network. It is based on the concept that connections from important nodes contribute more to the importance of a node than connections from less important nodes. Mathematically, \textit{PageRank} centrality can be defined as the eigenvector associated with the eigenvalue of 1 of the Google matrix $G$, given by the equation:
\begin{equation}
    p =G p =[\alpha S+ (1-\alpha) E] p
\end{equation}
where $S$ is the adjacency matrix of the network, with $1/N$ in the columns corresponding to dangling nodes,
$E$ is the matrix with value $1$ in all its $N\times N$ elements, and
$\alpha$ is a damping factor typically set to $\alpha=0.85$ \cite{leoGoogle}.

Finally, we introduce the \textit{maximal} and \textit{minimal} strategies based on the change in the giant component size when a route is removed from the network.
We define $\Delta(S_1)^t_i$ for node $i$ in $\mathbb{C}$-space as:
\begin{equation}
    \Delta (S_1)_i^t = S_1^t - (S_1)_i ^{t+1}
\end{equation}
Here, $S_1^t$ represents the number of nodes in the giant component at step $t$ of the removal procedure, and $(S_1)_i^{t+1}$ represents the number of remaining nodes in the giant component at step $t+1$ if node $i$ is removed. Therefore, $\Delta(S_1)_i^t$ accounts for the importance of node $i$ in $\mathbb{C}$-space and measures the change in $S_1$. The \textit{maximal} strategy selects the node with the highest $\Delta(S_1)_i^t$, while the \textit{minimal} strategy selects the node with the lowest one.

\subsection{Synthetic PTN model}

In this section, we explain the null model introduced in \cite{VonFerber2007} for creating synthetic PTN. The model also described in \cite{VonFerber2009} effectively captures many global key properties of public transport networks by growing networks of attractive self-avoiding walks (SAW).

The PTN model comprises $R$ routes, each featuring $S$ stations. It is constructed on an $X \times X$ square lattice, where the vertices ($x$) and edges represent potential stations and routes, respectively. 
The construction of routes follows the recipe:
\begin{enumerate}
\item Construct the first route as a self-avoiding walk (SAW) of $S$ lattice sites.
    \item Construct the $R-1$ subsequent routes as SAWs 
    with the following preferential attachment rules
    \begin{enumerate}
        \item choose a terminal station at $x_0$ with 
        probability $p\sim k_{x_0} + a/X^2$
        \item choose any subsequent station $x$ of the 
        route with probability $p\sim k_x + b$
    \end{enumerate}
\end{enumerate}

\begin{figure}[H]
    \centering
    \includegraphics[width=1.\textwidth]{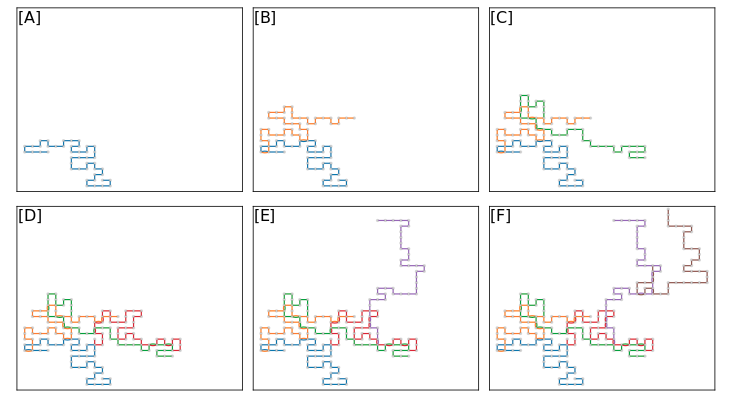}
    \caption{Illustration of the progressive process of synthetic networks generation for $a = 0$, $b = 5$, $S = 45$ and $X = 300$. In panel [A] a single route is shown and, as on the whole process, it is clearly a self-avoiding path. In panels [B], [C], [D], [E] and [F] a new route is subsequently added with a different color for clarity purpose. It is important to notice that there is a single connected component because the parameter $a$ is set to $0$.}
    \label{fig3}
\end{figure}

Here, $k_x$ represents the number of times the lattice site 
$x$ has been visited before (i.e., the number of routes that 
pass through $x$). To maintain the self-avoiding walk (SAW) 
property, any route that intersects itself is discarded. Figure \ref{fig3} illustrates the step-by-step process of building the synthetic model for a particular set of parameters. Each new route is added in a different color. 

In our study, we opt for closed boundary conditions instead of periodic boundary conditions, as the former aligns more closely with real-world metropolitan scenarios. We set the size of the lattice to $X=300$.

In the PTN model, we fix $a = 0$ and $b = 5$ to generate networks with properties similar to those of different cities, as done in \cite{VonFerber2009}. The generated network initially has only one component ($S_{\Omega_S}=1$ and $S_1=N$) for $a=0$, allowing a certain degree of randomness when routes pass through nodes not visited previously. For these parameter values, we apply the proposed attack methodology for various values of $R$ and $S$, which leads to different values of $N$.

\section{Synthetic PTN attack results} \label{results}

\begin{figure}[H]
    \centering
    \includegraphics[width=1.\textwidth]{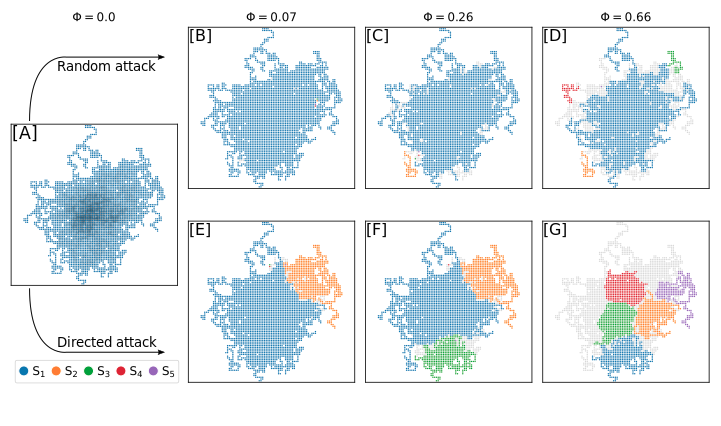}
    \caption{Illustration of the results of attacking a particular realization of a network generated with the synthetic model with $a=0$, $b=0.5$, $R=500$, $S=45$ a through a directed attack (using \textit{betweenness} centrality strategy) and a random attack. The original $\mathbb{L}$’-space is plotted on the panel [A]. Panels [B], [C], and [D] represent the first components resulting from different stages of the network random attack, with the corresponding fraction $\Phi$ of removed routes. Panels [E], [F], and [G] represent the results of the network- \textit{betweenness} directed attack for the same $\Phi$ values. Different colors represent the first five components into which the network is disassembled. Isolated nodes are plotted in gray.}
    \label{fig4}
\end{figure}

We systematically analyze multiple realizations of the synthetic network to explore the robustness of PTNs against attacks using the route removal  described procedure. Figure \ref{fig4} illustrates the procedure for a specific realization of the synthetic PTN, elucidating the analysis through a simple example. In the figure, the $\mathbb{L}$'-space of one realization of the synthetic model is depicted for $R = 500$ and $S = 45$ in the left panel ($\Phi=0$).
In the top row of panels, snapshots display different instances when routes are randomly removed, while in the bottom row, routes are removed with priority based on \textit{betweenness} in the $\mathbb{C}$-space.
The same number of removed routes is visualized for both processes, indicated by $\Phi=0.07$, $0.26$, and $0.66$ in the left-to-right columns, respectively. Each of the first five largest components is shown in different colors, as indicated in the legend, while the rest of the components ($S_i$ with $i>5$) are displayed in light gray.
Qualitatively, it is evident that targeted attacks following the \textit{betweenness} measure are more effective in detaching nodes from the first giant component. The second component appears with a size comparable to the giant component. Intimately related to this, it is also noticeable that the network disassembly mechanism, in the case of betweenness, leads to detachments of entire network regions, illustrated by the successive emergence of different components of considerable size (bottom right panel).

\begin{figure}[H]
    \centering
    \includegraphics[width=1.\textwidth]{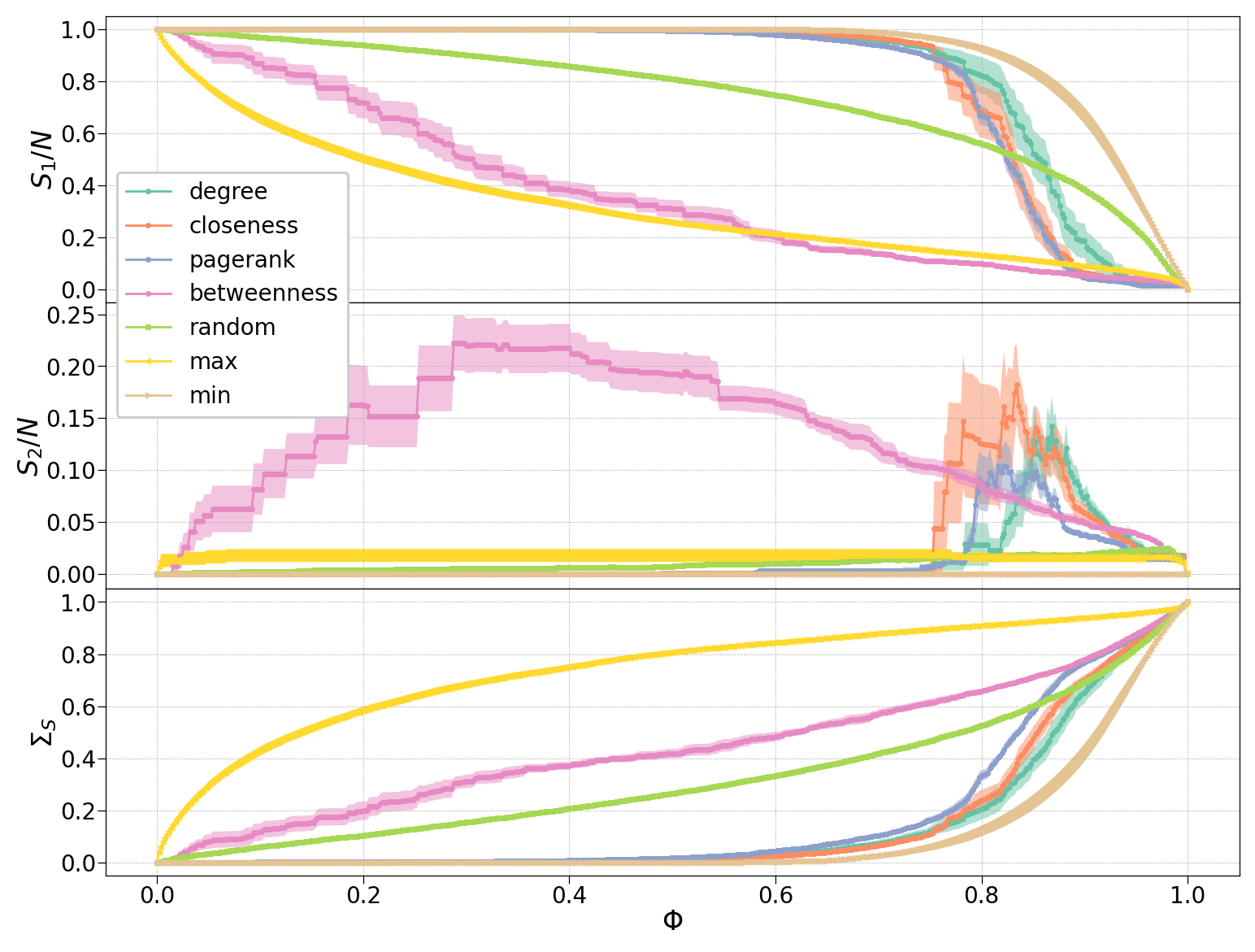}
    \caption{Average fraction of nodes on the two main components $S_1$ and $S_2$, and the normalized entropy $\Sigma_S$ are depicted as functions of $\Phi$ in the top, middle, and bottom panels, respectively. The estimation of the averages and their corresponding errors are taken over $10$ realizations of the synthetic network. The shaded areas correspond to the errors associated with the estimation of the averages. The different proposed removal processes are represented by colored lines for the synthetic PTN network with $a = 0$, $b = 5$, $S = 45$, and $R = 500$. "Max" and "Min" account for $maximal$ and $minimal$ strategies, respectively. This holds for next figures also.}
    \label{fig5}
\end{figure}

Going further, we will systematically analyze the attack following the procedures described in the previous section: four centrality measures in the $\mathbb{C}$-space (\textit{degree}, \textit{closeness}, \textit{PageRank} and \textit{betweenness}), the \textit{maximal} and \textit{minimal} removal procedure described in the previous section, and random removal. Each procedure is computed for $10$ realizations of the synthetic network model. For random removal, the number of realizations is increased to $100$, ensuring the convergence of the procedure. In this manner, we were able to reconstruct the average values of $S_1/N$, $S_2/N$, and $\Sigma_S$ as a function of $\Phi$, as shown in Figure \ref{fig5}, across the different realizations.

For $S_1/N$, the \textit{maximal} and \textit{minimal} procedures serve as lower and upper bounds for a wide range of $\Phi$. The \textit{betweenness} attack emerges as the most effective among the traditional centrality measures, leading to a substantial decrease in the giant component. Its effectiveness stems from its ability to break the network into several sizable components swiftly.  In contrast, the \textit{maximal} metric destructs the network similarly to \textit{betweenness} (or even more effectively for small $\Phi$) based on the giant component, though without generating the emergence of a second component. When considering the entropy, the distinction becomes more evident. The \textit{maximal} attack reveals its strategy: it removes routes to generate many isolated nodes, rapidly driving the entropy to high values. The disparity between random removal and classical metrics is notable. In the literature, for instance, as seen in \cite{Albert2000}, random attacks tend to be less detrimental when disassembling a network. 
The other three centrality measures in $\mathbb{C}$-space: \textit{degree}, \textit{closeness} and \textit{PageRank} exhibit similar behavior in $S_1$, $S_2$, and $\Sigma_S$.
One noticeable result is that the \textit{betweenness} procedure leads to smaller values of $S_1$ (for $\Phi>0.6$) compared to the \textit{maximal} method, which is defined to be the most effective one-step strategy.

\begin{figure}[H]
    \centering
    \includegraphics[width=1.\textwidth]{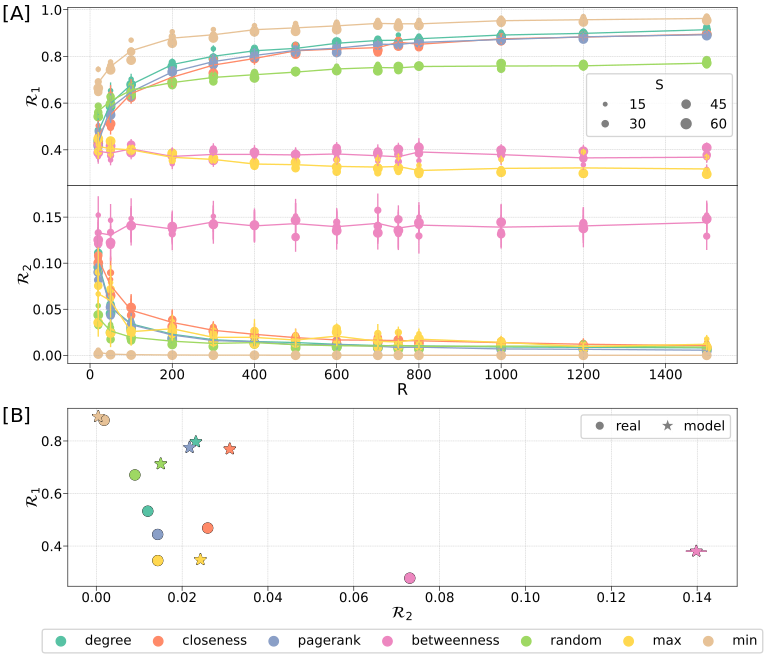}
    \caption{In panel [A], global robustness $\mathcal{R}_1$ y $\mathcal{R}_2$ as a function of the number of routes $R$, with their respective errors, inferred from $S_1$ and $S_2$. For each $R$, four different numbers of route stops, $S$, are represented by circle sizes. In addition, on panel [B], the relation between $\mathcal{R}_1$ and $\mathcal{R}_2$ is plotted for both the model (averaging over $R$ and $S$) and for the real case of Section \ref{real PTN}. Circles account for the actual network results; the average synthetic results are shown with stars and error bars represent the deviation of the average estimations, i.e. the average of the corresponding deviations.}
    \label{fig6}
\end{figure}

To ensure the robustness of the observed results against parameter choices, we compute $\mathcal{R}_1$ and $\mathcal{R}_2$ for synthetic networks with different values of $R$ and $S$, as illustrated in Figure \ref{fig6}. The behavior across different values of $S$, denoted by the varying size of the circles, demonstrates minimal variations. $\mathcal{R}_1$ and $\mathcal{R}_2$ stabilize as functions of $R$ for $R\gtrsim 200$. Initially, we observe that the procedures can be ordered by decreasing value of $\mathcal{R}_1$ as follows: \textit{minimal}; followed by the group of \textit{degree}, \textit{PageRank}, and \textit{closeness}; then random; and finally, the pair of \textit{maximal} and \textit{betweenness} with the lowest values. Notably, the attacks by \textit{betweenness} and \textit{maximal} exhibit distinct behavior from the rest, with \textit{maximal} being the most effective.

Regarding the second component, the order in $\mathcal{R}_2$ is as follows: \textit{betweenness} has the largest value by a considerable margin, followed by the group of \textit{closeness}, \textit{PageRank}, \textit{maximal}, and \textit{degree}; then random; and finally, the \textit{minimal} procedure.
It is noticeable that the \textit{maximal} procedure, which is the one with the best attack performance, together with \textit{betweenness}, does not yield high values of $\mathcal{R}_2$, but quite the opposite. This emphasizes that while both attack strategies have the same impact, for the case of the \textit{maximal} procedure, $S_1$ is not broken into large $S_2$. The bottom panel of Figure \ref{fig6} illustrates both results in the $\mathcal{R}_1$ and $\mathcal{R}_2$ plane.

\section{Real Network Analysis} \label{real PTN}

This section implements all defined attack procedures in a real PTN.
In particular, we will work with the urban bus system of the Metropolitan Area of Buenos Aires (AMBA), the largest metropolitan region in Argentina (and one of the largest in Latin America).

We have obtained the information  regarding the routes of the different $R=489$ bus lines operating in the AMBA using OpenStreetMaps \cite{OpenStreetMap}. By utilizing the DBSCAN algorithm implemented in \cite{pedregosa2011scikit}, nearby stops within an $80$ meter radius were merged, aiming to consider nearby stops as the same node. This process resulted in obtaining the $\mathbb{L}$'-space, which consists of $N=6016$ nodes and $22698$ links, with $\Omega_S=1$. As for the $\mathbb{C}$-space, the network comprises $R=489$ nodes and $19201$ links.

\begin{figure}[H]
    \centering
    \includegraphics[width=1.\textwidth]{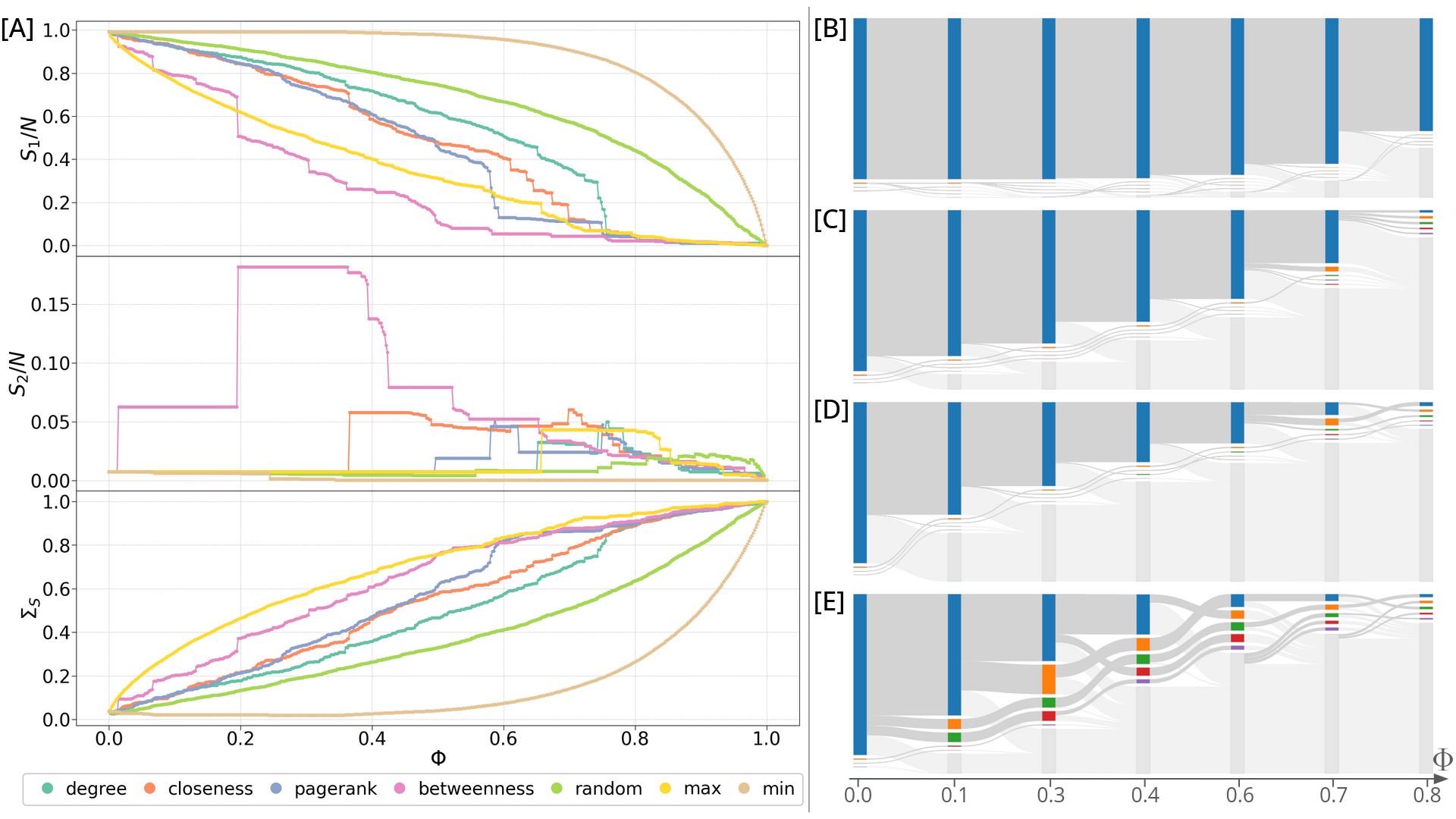}
    \caption{Left panel [A] $S_1/N$, $S_2/N$ and $\Sigma_S$ as a function of  $\Phi$ for the AMBA bus network. Right panel, Sankey diagrams for four different applied metrics: [B] \textit{minimal}, [C] \textit{degree}, [D] \textit{maximal} and [E] \textit{betweenness}. The flow between components at different $\Phi$ are plotted.}
    \label{fig7}
\end{figure}

The same analysis as in Section \ref{results} is conducted on the AMBA bus network. In panel $[A]$ of Figure \ref{fig7}, $S_1/N$, $S_2/N$, and $\Sigma_\mathcal{S}$ are plotted for the different attack strategies.

One notable observation is the significant difference compared to the results for the synthetic networks, as all metrics except the \textit{minimal} one are more effective at disrupting the network than the random strategy. This aligns with expectations from the literature and could indicate some shortcomings of the synthetic network model in mimicking real networks. In this real case, the \textit{minimal} method is the least effective in damaging $S_1$, while the \textit{betweenness} procedure emerges as the most effective for $\Phi>0.2$, surpassing the \textit{maximal} strategy.
In this case, both $S_1$ and $S_2$ exhibit considerable jumps, indicating a significant change when one route is attacked. It's important to note that the only way $S_2$ can increase is by forming a new and larger second component from the giant component.

Observing the evolution of $S_2/N$ reveals that \textit{betweenness} succeeds in disassembling the network by generating detachments of the giant component from important sets of connected nodes. Analogously, this behavior can be seen in the flow diagrams in panels $[B]$, $[C]$, $[D]$, and $[E]$, where the flow of nodes between the first five largest components is plotted for the \textit{minimal}, \textit{degree}, \textit{maximal}, and \textit{betweenness} strategies, respectively.
For example, the \textit{minimal} and \textit{degree} attacks disassemble the network by removing nodes that are then isolated  throughout the process. The same happens, but more quickly, for the \textit{maximal} case. In contrast, the \textit{betweenness} strategy quickly generates the detachment of secondary components that prevail throughout the attack process.

Interestingly, when observing the normalized entropy associated with the sizes of the different components throughout the process, the \textit{minimal} and \textit{maximal} strategies represent the lower and upper bounds. Although \textit{betweenness} is more effective in breaking the network, the distribution of components along the process exhibits a lower entropy than for the \textit{maximal} case. Additionally, we again observe a discrepancy with the synthetic network: the entropy for the random case is lower than that for the classical metrics. This difference highlights a key distinction between synthetic and real networks.

To quantitatively compare the overall robustness against the different strategies and how the network is disassembled during the process, we compute $\mathcal{R}_1$ and $\mathcal{R}_2$ for the actual network in the bottom panel of Figure \ref{fig6}. Thus, \textit{betweenness} stands out as the most effective strategy due to its low $\mathcal{R}_1$, while it has the highest $\mathcal{R}_2$.

The strategies applied to the AMBA PTN can be classified in the $\mathcal{R}_1$ and $\mathcal{R}_2$ plane, where \textit{betweenness} is always located in the bottom right corner and \textit{minimal} in the opposite one. The results of the real PTN (shown in circles) can be compared with the cases of synthetic modeled PTN (represented by star symbols with error bars denoting the standard deviation obtained from $10$ realizations). Although they do not perfectly coincide, there is a similar behavior between the strategies applied to the real PTN and the modeled one.

\begin{figure}
    \centering
    \includegraphics[width=.7\textwidth]{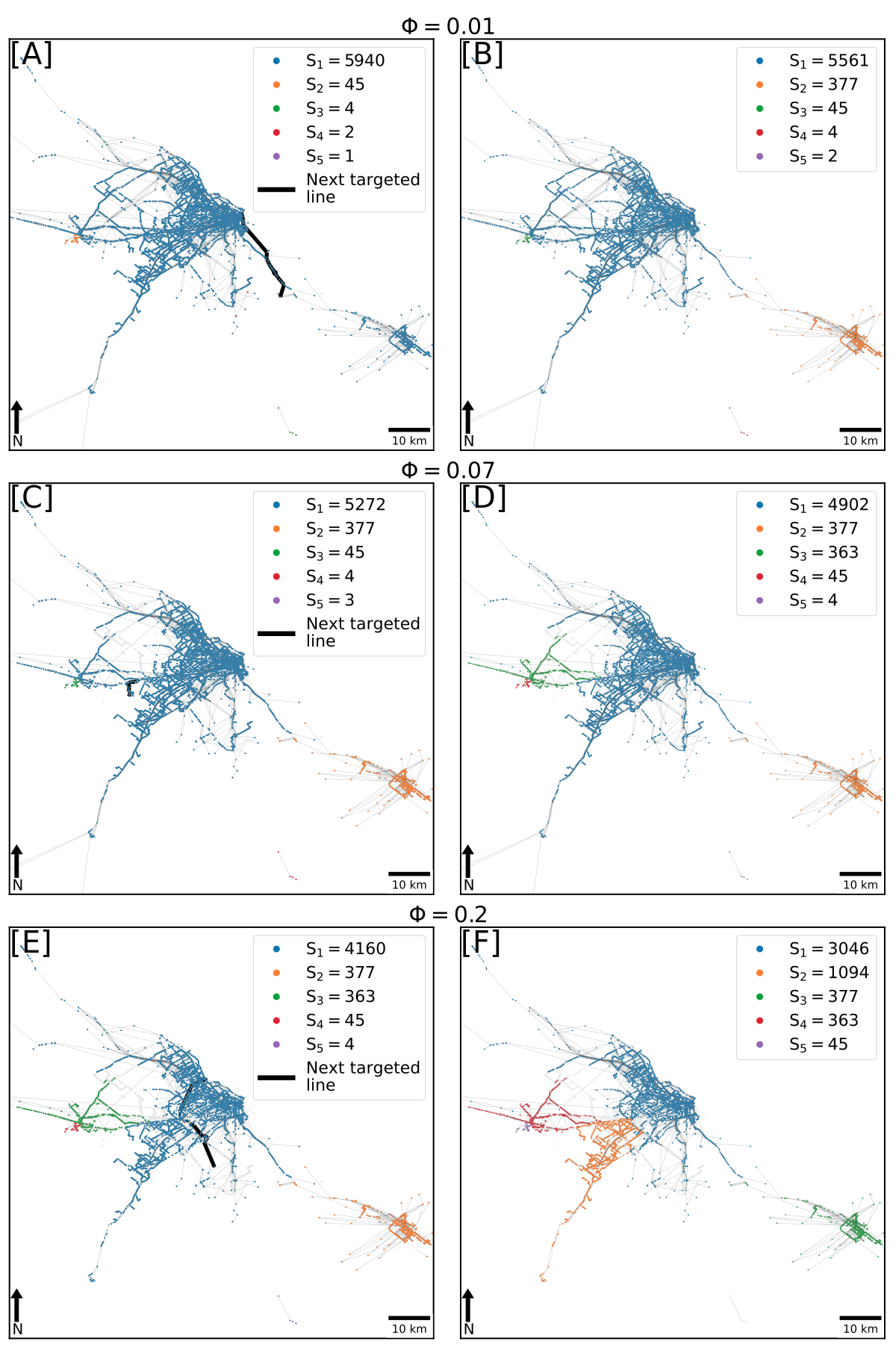}
    \caption{Three stages of the directed attack process on the $\mathbb{L}$’-space of the AMBA region, using the \textit{betweenness} strategy. Each row accounts for a different fraction $\Phi$ of removed routes and visualizes the effect of removing central routes. On panels [A], [C], and [E], the route with the higher \textit{betweenness} is highlighted in black. On panels [B], [D], and [F], the network after the removal of the targeted lines is shown. Node colors represent the component it belongs to, showing the effects of these attacks on the first five components of the network and their sizes on the plot legends.}
    \label{fig8}
\end{figure}

To conclude the analysis of the real network and better understand the disassembly process of the network, we decided to study the \textit{betweenness} removal strategy for different values of $\Phi$. In particular, we wish to examine what happens when there are jumps in $S_2/N$ in Figure \ref{fig7}[A]. For this purpose, in Figure \ref{fig8}, we plot the $\mathbb L$'-space of the AMBA for 3 different values of $\Phi$. In the upper panel, we mark the line to be removed in the next step, while in the lower panel, we observe the effect produced by such removal. The different components are plotted in different colors.

The first observation is that removing the path with the highest \textit{betweenness} does not necessarily result in the appearance of a new second giant component. For instance, at $\Phi = 0.07$, $S_2$ remains constant, while a new $S_3$ emerges. However, what appears to be more significant is that in all three cases, the components detaching from the lattice belong to the periphery of the lattice. A similar effect is observed in the example shown in Figure \ref{fig4}. This suggests that the routes with higher \textit{betweenness} in the $\mathbb C$-space serve as bridges between the core of the network and its expansions into the periphery.

We found that the results obtained from the model were not significantly different from those of the real network. However, to understand the differences in $S_1$ and $S_2$ between them, we can propose certain improvements to the model. One major simplification of the synthetic model is assuming all routes ($R$) have the same number of stations ($S$). However, this assumption does not hold in the real network, where routes have a varied number of stations. Therefore, by considering the same synthetic model but with a distribution of stations on the routes forced to be similar to the real one, we obtain results that are much more similar to those of the real network.
For example, in cases of $S_1(\Phi)$, we can observe how the procedures given by the metrics in the $\mathbb{C}$-space (\textit{degree}, \textit{closeness}, and \textit{PageRank}) are more harmful than the random ones in these synthetic models with "realistic" routes. This indicates that improvements can be proposed to the synthetic model to mimic the topological and robustness properties against attacks more accurately, which goes beyond the objectives of this work.

\section{Discussion} \label{discussion}

In this work, we focused on studying the robustness and sensitivity of public transport networks (PTNs) to route attacks. 
This novel approach allows for obtaining different results than node or link attacks on the network. 
We defined various network attack procedures based on centrality measures for $\mathbb{C}$-space, where nodes represent routes.
As metrics of damage, we used measures based on the sizes of the components formed in the networks where stations are nodes (both $\mathbb{L}$-space and $\mathbb{L}'$-space). 
Additionally, we introduced two new route attack procedures, \textit{maximal} and \textit{minimal}, based on the most and least harmful tactics in the network damage at the next step. 
All of this was studied in both synthetic PTNs, which allowed for statistical analysis, and a real network of considerable size and well-established structure in Buenos Aires. 
We observed how different centrality measures or strategies generated distinct outcomes. 
Specifically, we illustrated that \textit{degree}, \textit{closeness}, and \textit{PageRank} metrics show similar behaviors while \textit{betweenness} operates differently.
The latter procedure progressively breaks down the network into several components of comparable sizes. 
We found that the \textit{minimal} strategy inflicts the  most minor damage to the network. 
However, following the \textit{betweenness} method can either surpass or be on par with the \textit{maximal} strategy in terms of damage, albeit with markedly different behavior in the second component. 
Finally, we propose improvements to synthetic networks by observing the differences in the behavior between the random strategy and the other three centrality measures across synthetic and real networks.
These enhancements in properties may contribute to producing new synthetic models of metropolitan areas for future studies. 
We also believe that the analysis developed in this work can aid in enhancing and designing PTNs that are efficient and resilient to difficulties. Furthermore, we believe that continuing in this direction would make it feasible to assess the accessibility and efficiency of a PTN by utilizing actual data on residents' mobility patterns.

\section*{Funding}
We acknowledge the support of PIP 2021-2023 11220200102701CO from CONICET for their financial assistance.

\section*{Declaration of Competing Interest}
The authors declare that they have no known competing financial interests or personal relationships that could have appeared to influence the work reported in this paper.
\section*{Contributions}
TC collected and analyzed the data. TC, IC and LE interpret the results and wrote the manuscript. EL and IC conceived the study. All authors read and approved the final manuscript.



\bibliographystyle{ieeetr} 

\end{document}